\documentclass[prl,twocolumn,showpacs,superscriptaddress,equation]{revtex4-1}
 \pdfoutput=1
\usepackage{times}
\usepackage{graphicx}
\usepackage{amsmath}

\begin{document}
\title{Investigating Atomic Contrast in Atomic Force Microscopy and Kelvin Probe Force Microscopy on Ionic Systems using Functionalized Tips}

\author{Leo \surname{Gross}}
\email{lgr@zurich.ibm.com}
\affiliation{IBM Research -- Zurich, 8803 R\"uschlikon,
Switzerland}

\author{Bruno \surname{Schuler}}
\affiliation{IBM Research -- Zurich, 8803 R\"uschlikon, Switzerland}
\author{Fabian \surname{Mohn}}
\affiliation{IBM Research -- Zurich, 8803 R\"uschlikon, Switzerland}
\affiliation{Present address: ABB Corporate Research, 5405 Baden-D\"attwil, Switzerland}
\author{Nikolaj \surname{Moll}}
\affiliation{IBM Research -- Zurich, 8803 R\"uschlikon, Switzerland}
\author{Niko \surname{Pavli\v{c}ek}}
\affiliation{IBM Research -- Zurich, 8803 R\"uschlikon, Switzerland}
\author{Wolfram \surname{Steurer}}
\affiliation{IBM Research -- Zurich, 8803 R\"uschlikon, Switzerland}
\author{Ivan \surname{Scivetti}}
\author{Konstantinos \surname{Kotsis}} 
\author{Mats \surname{Persson}}
\affiliation{Surface Science Research Centre, Department of Chemistry, University of Liverpool, Liverpool, L69 3BX, United Kingdom}
\author{Gerhard \surname{Meyer}}
\affiliation{IBM Research -- Zurich, 8803 R\"uschlikon, Switzerland}

\date{\today}

\begin{abstract}
Noncontact atomic force microscopy (NC-AFM) and Kelvin probe force microscopy (KPFM) have become important tools for nanotechnology; however, their contrast mechanisms on the atomic scale are not entirely understood. Here we used chlorine vacancies in NaCl bilayers on Cu(111) as a model system to investigate atomic contrast as a function of applied voltage, tip height, and tip functionalization. We demonstrate that the AFM contrast on the atomic scale decisively depends on both the tip termination and the sample voltage. On the contrary, the local contact potential difference (LCPD) acquired with KPFM showed the same qualitative contrast for all tip terminations investigated, which resembled the contrast of the electric field of the sample. We find that the AFM contrast stems mainly from electrostatic interactions but its tip dependence cannot be explained by the tip dipole alone. With the aid of a simple electrostatic model and by density functional theory we investigate the underlying contrast mechanisms. 
\end{abstract}


\maketitle

\section{I. Introduction}
Atomic force microscopy becomes increasingly important for studying surfaces on the atomic scale~\cite{Giessibl2003,Morita2009,Barth2011} as it provides atomic resolution and is not restricted to conducting samples. Furthermore, a wealth of information can be obtained by spectroscopic methods. For example, atomic resolution with chemical sensitivity was demonstrated using force-distance spectroscopy~\cite{Sugimoto2007a}. Using the force-voltage spectroscopy mode for Kelvin probe force microscopy (KPFM), charge states of single atoms~\cite{Gross2009}, defects~\cite{Konig2009}, and molecules~\cite{Leoni2011} were determined, and submolecular resolution was obtained~\cite{Mohn2012,Kawai2013,Schuler2014}. The atomic contrast observed with KPFM~\cite{Bocquet2008,Enevoldsen2008,Sadewasser2009,Yurtsever2012,Ma2013} triggered efforts to explain the underlying contrast mechanism theoretically~\cite{Nony2009,Bocquet2011,Masago2010,hynninen2011,Sadeghi2012}. The most important open questions are: What are the physical properties mapped by AFM and KPFM on the atomic scale, and how can we take advantage of this information?

The (100) surfaces of alkali halides are often used as model systems to investigate atomic contrast on insulators by NC-AFM~\cite{Kantorovich2000,Giessibl1992,Hoffmann2004,Lantz2006,Foster2009,Ruschmeier2008,Fremy2012}, and theory predicted that the polarity of the tip apex determines whether the largest attractive forces are measured above anions or cations~\cite{Hofer2003}. A direct way to identify the lattice sites experimentally, without a priori knowledge of the tip termination and its imaging contrast, is the application of markers with known adsorption sites~\cite{teobaldi2011}. We used Cl vacancies in the top layer of NaCl(100), which have previously been studied using scanning tunneling microscopy (STM) by Repp et al.~\cite{Repp2005a}, to provide unambiguous lattice site identification. Moreover, this vacancy is an atomically well-defined, highly symmetric defect and thus it can be used to verify atomic resolution and to identify tip asymmetries. Finally, the Cl vacancy is also useful to investigate the spatial resolution of KPFM as it provides an uncompensated positive charge within the ionic lattice~\cite{Repp2005a}. Here we used Cl vacancies in a bilayer of NaCl(100) on Cu(111) as model systems to study the atomic contrast of AFM and KPFM on ionic systems employing four different tips terminated with individual Cu, Au, Cl, and Xe atoms. We demonstrate atomic resolution with AFM and KPFM for all four tip functionalizations investigated. We found that the AFM contrast does depend crucially on the sample bias and the tip termination, whereas the local contact potential difference (LCPD) does not. The AFM contrast arises mainly from electrostatic interactions but it cannot always be explained by the tip polarity.

\section{II. Experiment}
The measurements were performed at a sample temperature of $T$\,=\,5\,K with a combined low-temperature STM/AFM based on a qPlus tuning fork sensor design~\cite{Giessibl1999} operated in the frequency modulation mode~\cite{Albrecht1991}. We grew two-monolayer (ML) thick (100)-oriented NaCl islands on Cu(111), denoted as NaCl(2\,ML)/Cu(111), and adsorbed Au and Xe atoms at $T$\,$<$\,10\,K on the sample~\cite{Mohn2013}. As tip we used a PtIr wire cut by a focused ion beam. The tip was repeatedly indented into the Cu(111) substrate to form a Cu-terminated tip (Cu tip). Starting from a Cu tip, we fabricated an Au-terminated tip (Au tip) by picking up individual Au adatoms from NaCl(2\,ML)/Cu(111)~\cite{Repp2004}. We used vertical manipulation to create the Cl-vacancies and confirmed their formation by scanning tunneling spectroscopy~\cite{Repp2005a}. To this end, a non-oscillating Cu tip was approached by about 5\,$\text{\AA}$ at zero sample voltage ($V$), starting from a tunneling set point of $V$\,=\,200\,mV, $I$\,=\,2\,pA above NaCl(2\,ML)/Cu(111). This approach often resulted in the transfer of a Cl atom from the sample towards the tip, forming a Cl-terminated tip (Cl tip) and creating a Cl vacancy at the surface. An Xe-terminated tip~\cite{Eigler1991,Mohn2013} (Xe tip) was created by picking up Xe atoms with a Cu tip from third-layer NaCl step edges.

We calibrated the offset of the tip height such that for each tip, the tip height $z$\,=\,0\,$\text{\AA}$ corresponds to the smallest tip height measured. When we further decreased $z$, we usually observed instabilities in the ${\Delta}f$ signal, indicating atomic rearrangements at the tip-sample junction [one such tip change can be observed in Fig.\,3(e)]. Note that for each tip there is a different unknown offset of $z$ with respect to the separation of the last atom of the tip and the sample surface. We set $z$ by applying a height offset ${\Delta z}$ with respect to the tunneling setpoint ($V_{\text{sp}}$, $I_{\text{sp}}$) above NaCl(2\,ML)/Cu(111). For the tips shown $z$\,=\,0\,{\AA} corresponded to ($V_{\text{sp}}$, $I_{\text{sp}}$, ${\Delta z}$) = (100\,mV, 100\,pA, --0.9\,\AA) for the Cu tip, (200\,mV, 100\,pA, --1.6\,\AA) for the Au tip, (100\,mV, 30\,pA, --0.8\,\AA) for the Cl tip, and (200\,mV, 60\,pA, --0.3\,\AA) for the Xe tip, respectively. The Cu tip makes point contact with the surface, thereby creating vacancies, at about $z$\,=\,--2\,{\AA}. In this case we can make a rough estimate of the distance of the center of the Cu tip atom to the Cl surface anion to correspond to the sum of their van der Waals radii (i.e. 3.0\,\AA). Therefore $z$\,=\,0\,\AA\, corresponds to a distance of about 5\,\AA\,between the foremost Cu tip atom and surface. This is in good agreement with calculations (see below) from which we estimated that for all tips $z$\,=\,0\,$\text{\AA}$ corresponds to a distance of about 5\,\AA\ to 6\,\AA\ between the tip atom center and the top NaCl layer.

\begin{figure}
\begin{center}
\includegraphics[width=0.9\linewidth]{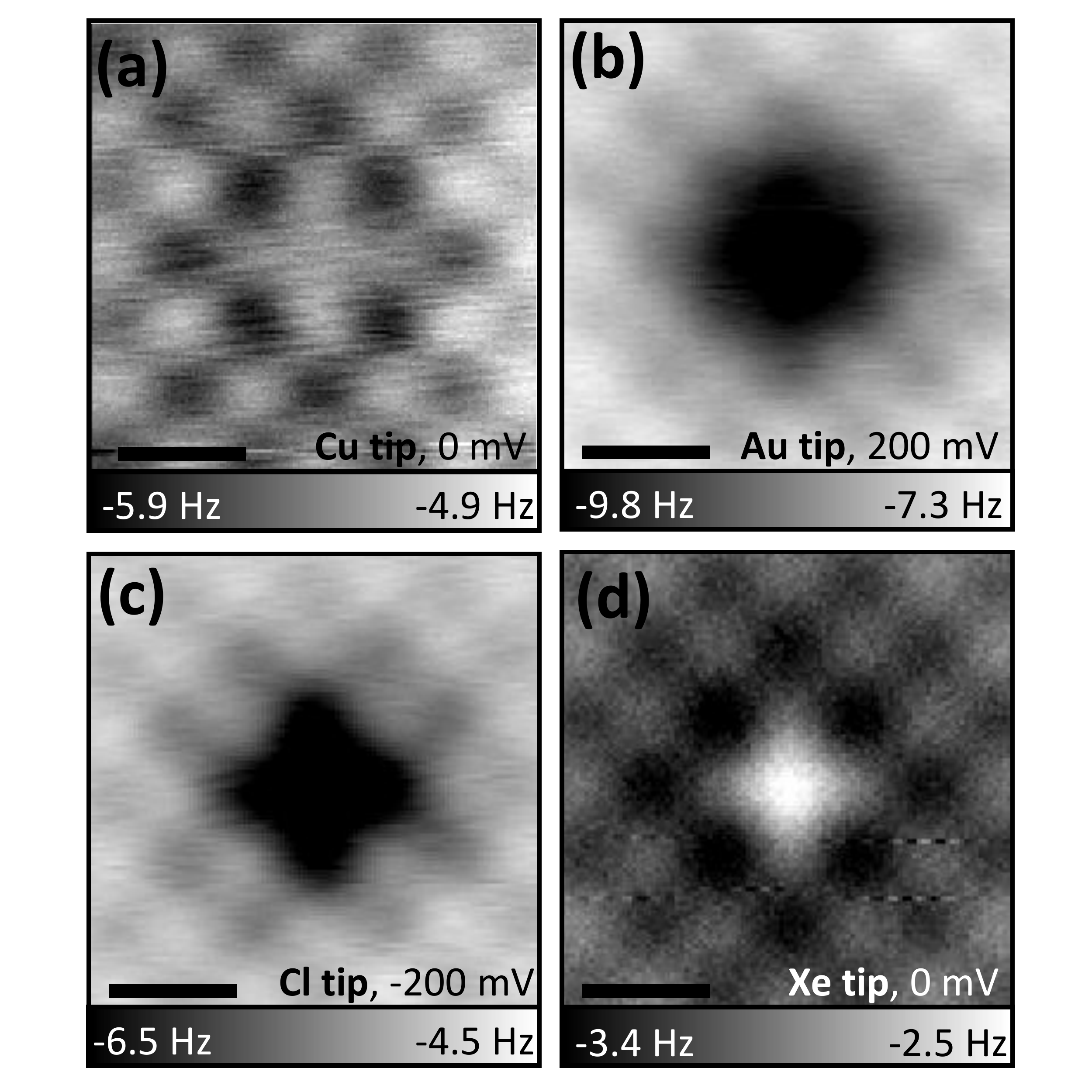}
\caption{\label{fig1}\small Constant-height AFM maps obtained at constant sample voltages $V$ acquired with tips terminated with Cu, Au, Cl and Xe, respectively. (a) Cu tip, $V$\,=\,0\,mV, (b) Au tip, $V$\,=\,200\,mV, (c) Cl tip, $V$\,=\,--200\,mV, (d) Xe tip, $V$\,=\,0\,mV. The voltage $V$ was chosen to roughly compensate for the LCPD.
}
\end{center}
\end{figure}

Figure~\ref{fig1} displays NC-AFM images of Cl vacancies in the top layer of bilayer NaCl on Cu(111) using four tips with different atomic terminations: Cu, Au, Cl and Xe. The images show $\Delta f$ acquired at constant height and constant sample voltage $V$. The voltages $V$ were chosen to roughly (within a few hundreds of mV) compensate for the respective LCPD as described below. The same four tips were used for all measurement shown in this paper (except Fig.\,A5). Note that the Cl vacancy unambiguously indicates the location of the Na and Cl sites. Obviously the tip termination plays an important role in the atomic contrast achieved with AFM.

\begin{figure}
\includegraphics[width=1\linewidth]{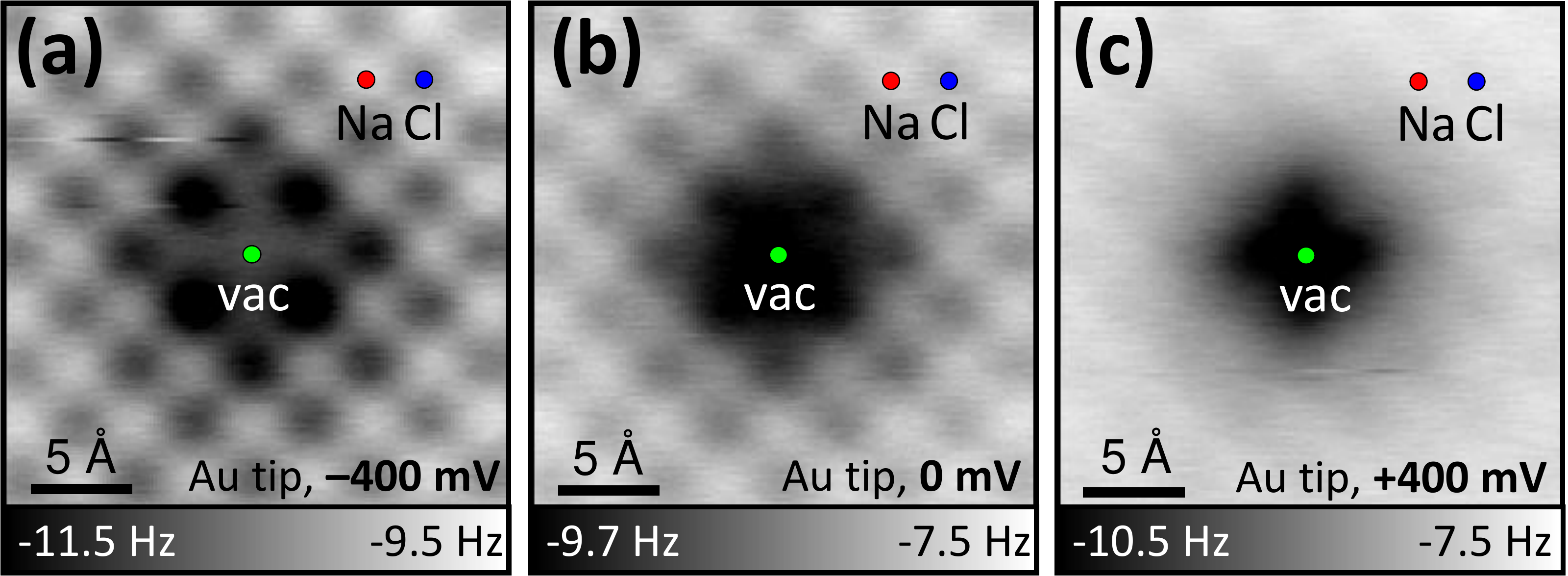}
\caption{\label{fig2}\small (Color online) Constant-height AFM measurements of a Cl vacancy in bilayer NaCl on Cu(111) using an Au tip at a sample voltage of (a) $-400\,$mV, (b) $0\,$mV, and (c) $+400\,$mV. Oscillation amplitude $A$\,=\,0.5\,$\text{\AA}$ and tip height $z$\,=\,0.0\,$\text{\AA}$. The positions of the Na site (Na, red), the Cl site (Cl, blue), and the vacancy site (vac, green) are indicated.
}
\end{figure}

AFM images at constant tip height $z$ with the Au-terminated tip (Au tip) at different sample voltages $V$ are shown in Fig.~\ref{fig2}. For all voltages shown the Cl sites exhibit a smaller (more negative) frequency shift ${\Delta}f$ than the Na sites. This difference corresponds to a larger attraction above the Cl sites than above the Na sites, in agreement with an investigation by Teobaldi et al.~using metal tips~\cite{teobaldi2011}. However, we observed that the atomic contrast crucially depended on $V$: With increasing $V$, the contrast between Cl and Na sites decreased and the vacancy site appeared darker (more attractive) than the Na and Cl sites. At $V$\,=\,--400\,mV [Fig.~\ref{fig2}(a)], the sites exhibiting the smallest ${\Delta}f$ are the four neighboring Cl sites of the vacancy, whereas at $V$\,=\,+400\,mV [Fig.~\ref{fig2}(c)] the vacancy site itself and its four neighboring Na sites exhibit the smallest ${\Delta}f$. These $V$-dependent contrast inversions on the atomic scale demonstrate the importance of taking into account the $V$ dependence, in addition to the tip termination and the $z$ dependence.

\begin{figure*}
\includegraphics[width=0.9\linewidth]{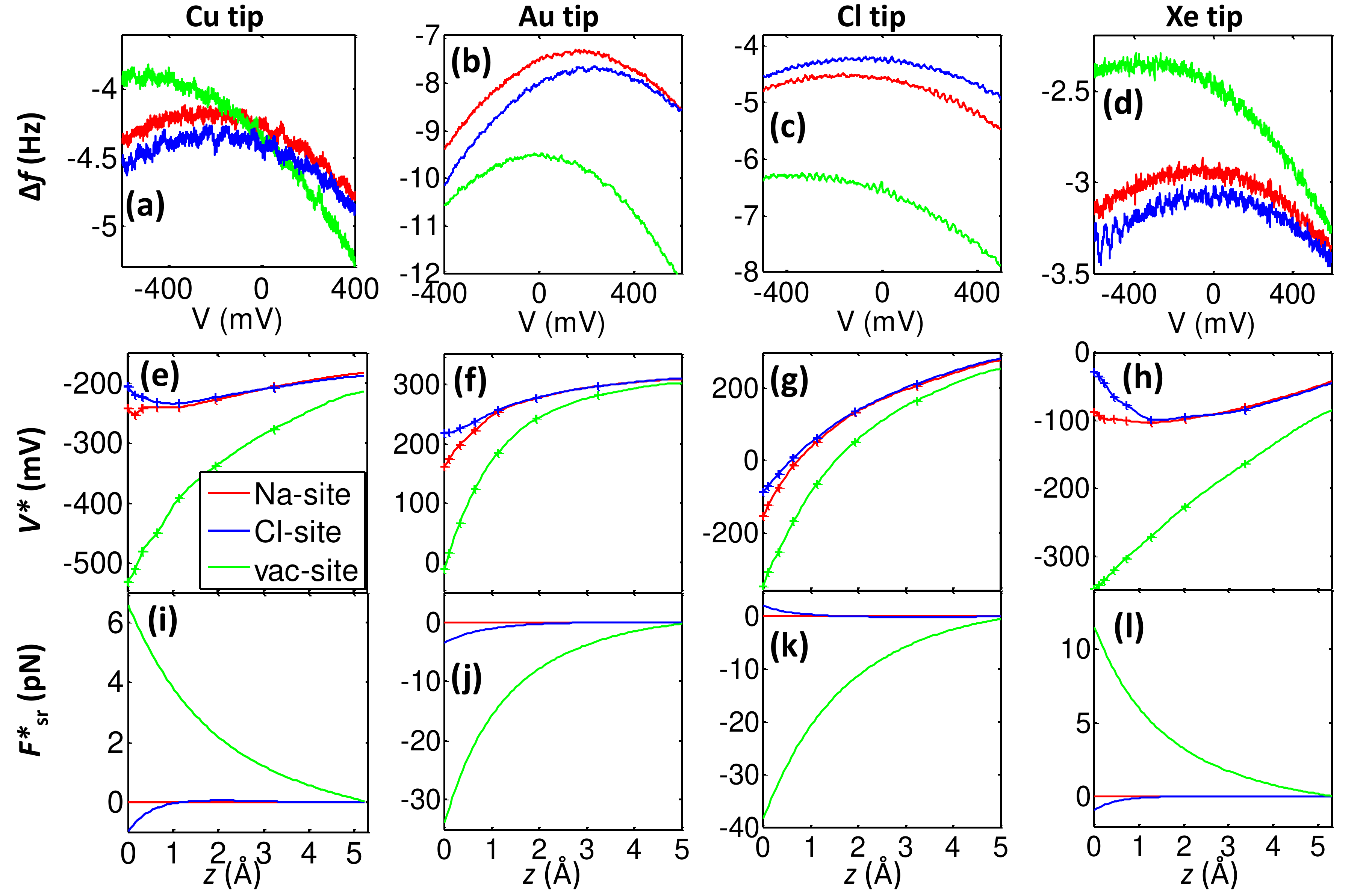}
\caption{\label{fig3}\small (Color online) Force spectroscopy with Cu tip (1st column), Au tip (2nd column), Cl tip (3rd column) and Xe tip (4th column) at the positions indicated in Fig.~\ref{fig2} of Na (red), Cl (blue) and the vacancy site (green). Panels (a)-(d): ${\Delta}f(V)$ at $z$\,=\,0\,\AA. Panels (e)-(h): $V^*(z)$; measured values are indicated by crosses, lines are interpolated. Panels (i)-(l): LCPD-compensated short-range forces $F^*_{\text{sr}}(z)$ with respect to Na sites.
}
\end{figure*}

We used force-voltage spectroscopy to characterize the different tips. We obtained ${\Delta}f(V)$ spectra with the Cu, Au, Cl and Xe tip above the atomic sites of the vacancy (vac), Na, and Cl, respectively, shown for $z$\,=\,0\,\AA\,in Fig.~\ref{fig3}(a)-(d). (The complete data sets, including different tip heights $z$, are shown in the appendix). We fitted each spectrum with a parabola and obtained the peak frequency shift ${\Delta}f^{*}$ and the peak voltage $V^{*}$. The latter corresponds to the LCPD divided by the elementary charge, and $\Delta f^{*}$ is the frequency shift at $V^{*}$. It is important to note that for different tips, even with the same atomic functionalization, we obtained $V^{*}$ that could be offset by several 100\,mV and we also observed quantitative differences in the $V^{*}$ contrast. However, the qualitative contrast of $V^{*}$ and $\Delta f^*$ were reproducible with the respective tip functionalizations. In general, the $V^{*}$ contrast increased for tips exhibiting small ${|{\Delta}f}|$ (i.e., presumably very sharp tips with small background forces) and we selected such tips for our experiments. 

Figure~\ref{fig3}(e)-(h) shows $V^{*}(z)$. For $z$\,$\gtrsim$\,2\,\text{\AA}, we observed that $V^{*}$ generally increased with increasing $z$. This we attributed to averaging effects caused by surfaces with different work functions in the junction: With increasing $z$, the contribution of the atomic junction relative to the contribution of the mesoscopic junction will decrease. The NaCl(2\,ML) islands were surrounded by Cu(111), which has a larger work function than NaCl, explaining the increase in $V^*$ for large $z$~\cite{Gross2009, Bieletzki2010}. Due to these averaging effects the quantitative $V^*$ (LCPD) contrast will also be a function of the tip size, as explained by Bieletzki et al.~\cite{Bieletzki2010}. For smaller tip heights (on the order of the atomic scale), the influence of the atomic tip termination and the surface atomic structure on $V^*$ will increase.  For $z$\,$\lesssim$\,1\,\text{\AA}, not only Na and Cl sites can be distinguished, but also the influence of the tip termination becomes apparent. For example, in the case of the Cl tip, the large positive slope of $V^{*}(z)$ above the Cl site even for small $z$ can be explained by a negatively charged Cl tip-atom.

For all tips investigated we obtained a similar LCPD contrast: ${V^{*}_{\text{Cl}} > V^{*}_{\text{Na}} \gg V^{*}_{\text{vac}}}$. For each tip, the difference at $z$\,=\,0\,\text{\AA} between the Na and Cl sites is several tens of mV, whereas the difference between the vacancy and NaCl is a few hundreds of mV. With increasing $z$, the $V^*$ contrast between Na and Cl sites decays faster than the contrast between the vacancy and NaCl, which can be explained by the long-range electrostatic potential of the uncompensated positive charge of the vacancy~\cite{Repp2005a}.

Next we investigated the differences in the LCPD-compensated short-range forces $F^*_{\text{sr}}(z)$ shown in Fig.~\ref{fig3}(i)-(l). To obtain $F^*_{\text{sr}}(z)$ from the ${\Delta}f(V)$ spectra measured at discrete tip heights $z$ we used a similar method as Sadewasser et al.~\cite{Sadewasser2009}. First we interpolated spectra to obtain ${\Delta}f(V, z)$ on a denser grid in $z$ than used in the experiments. To this end, the measured ${\Delta}f(z)$ at fixed $V$ were fitted by a spline fit. Next we determined the ${\Delta}f^{*}$ values of the ${\Delta}f(V)$ parabolas at fixed $z$ and obtained ${\Delta}f^{*}(z)$. Then we applied the Sader method~\cite{Sader2004} to ${\Delta}f^{*}(z)$, yielding the LCPD-compensated forces $F^{*}(z)$. To obtain the site-specific differences of the forces, $F^*_{\text{sr}}(z)$, and suppress background forces, we subtracted the forces measured above the Na site $F_{\text{Na}}^*(z)$ from $F^*(z)$. 

\begin{figure*}
\includegraphics[width=0.9\linewidth]{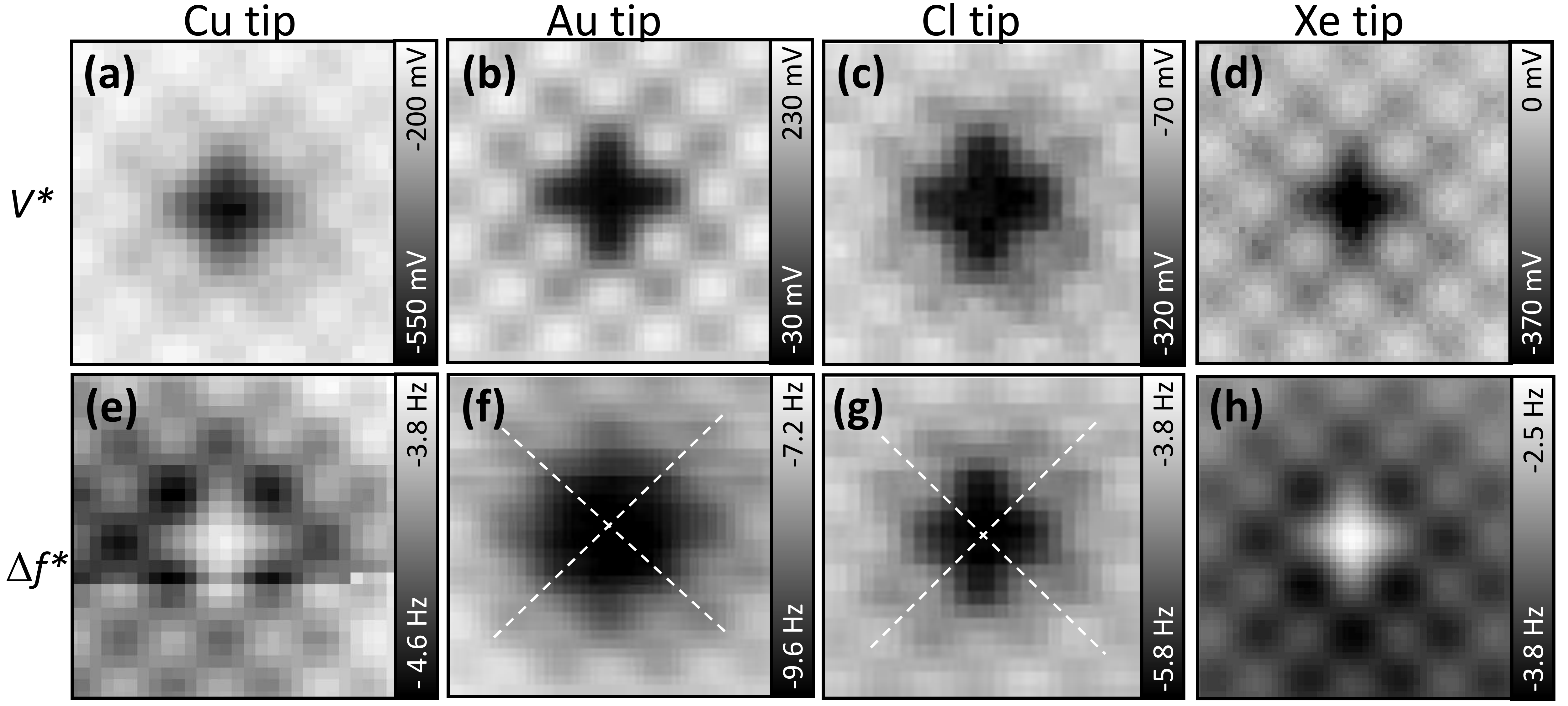}
\caption{\label{fig4}\small KPFM maps of $V^*$ (a)-(d) and $\Delta f^*$ (e)-(h) above a Cl vacancy using the same Cu, Au, Cl, and Xe tip as in Fig.~\ref{fig3}, respectively. All images are of size $18\,\text{\AA}\,\times\,18\,\text{\AA}$, with $A\,=\,0.5\,\text{\AA}$. The tip height is $z\,=\,0.0\,\text{\AA}$ for all images, except for (c) and (g), where $z\,=\,0.1\,\text{\AA}$. The dashed lines in (f) and (g) indicate lines of Cl sites. The contrast change in (e) is probably caused by a slight rearrangement of the Cu tip atoms. Note that this had little effect on the simultaneously measured $V^{*}$ map [see (a)].
} 
\end{figure*} 

At $z$\,=\,0\text\,{\AA}, the LCPD-compensated forces above the reference Na site, $F_{\text{Na}}^*($z$\,=\,0\,\text{\AA})$, were $-139\,$pN for the Cu tip, $-264\,$pN for the Au tip, $-126\,$pN for the Cl tip, and $-76\,$pN for the Xe tip. For all tips and sites investigated, $F^{*}(z)$ [and also $\Delta f^{*}(z)$] is negative and monotonically increasing with $z$, indicating that we always measured in the attractive regime of the interaction. For all tips the contrast of $F^*_{\text{sr}}$ decreased with increasing $z$. The difference between the $F^*_{\text{sr}}$ of the vacancy and of the Na and Cl sites was about one order of magnitude larger than the corresponding difference between Na and Cl sites. We observed no contrast inversions of $F^*_{\text{sr}}$  as a function of $z$. However, we observed different signs of $F^*_{\text{sr}}(z)$ for the different tips: The vacancy appeared more attractive than the Na and Cl sites with the Au and Cl tips, but less attractive with the Cu and Xe tips. The Cl site appeared more attractive than the Na site for the Cu, Au and Xe tips, but less attractive with the Cl tip.

To investigate the lateral contrast and resolution of KPFM, we measured maps of $V^{*}$ and ${\Delta f^*}$ by acquiring ${\Delta}f(V)$ spectra on a grid~\cite{Mohn2012}, shown in Fig.~\ref{fig4}. For all tips we obtained atomic resolution in both the $V^{*}$ and the ${\Delta f^*}$ channel. The $V^{*}$ maps of all tips [Fig.~\ref{fig4}(a)-(d)] appear very similar. Not only is the $V^{*}$ contrast of the three different sites similar, but also the shift of $V^{*}$ towards more negative values above the four neighboring Na sites of the vacancy is exhibited by all tips. As a result the vacancy appears in the shape of a dark cross in all $V^{*}$  maps. In contrast, the ${\Delta f^*}$ maps [Fig.~\ref{fig4}(e)-(h)] show qualitatively different contrasts for different tips.

These results indicate that the qualitative contrast of the $V^*$ (the LCPD) is in general tip-independent. Note that quantitatively the LCPD depends on the tip height and tip shape~\cite{Gross2009, Bieletzki2010, Bocquet2011} and can be expected to become qualitatively tip-dependent due to tip relaxations, e.g. for a CO tip at small tip height~\cite{Schuler2014}. The qualitative tip independence of the LCPD that we observe is a unique feature of KPFM, as both STM and AFM on the atomic scale show a qualitative dependence on the tip functionalization. This finding is in opposition to the conclusions of Yurtsever et al.\,who reported tip-dependent LCPD contrast inversion on CaF$_2$~\cite{Yurtsever2012}. Their interpretation was based on the assignment of the tip polarity by the contrast observed with AFM, rather then fabricating tips with known functionalization and using defects to select symmetric tips. However, as we demonstrate below for the Au tip, the AFM contrast cannot always be explained by the tip dipole. Our results indicate that in the case of an unknown tip termination, the LCPD contrast is of better use in the assignment of the lattice sites than the $\Delta f^*$ contrast. 

\section{III. Theory and Discussion}

\begin{figure}
\includegraphics[width=1\linewidth]{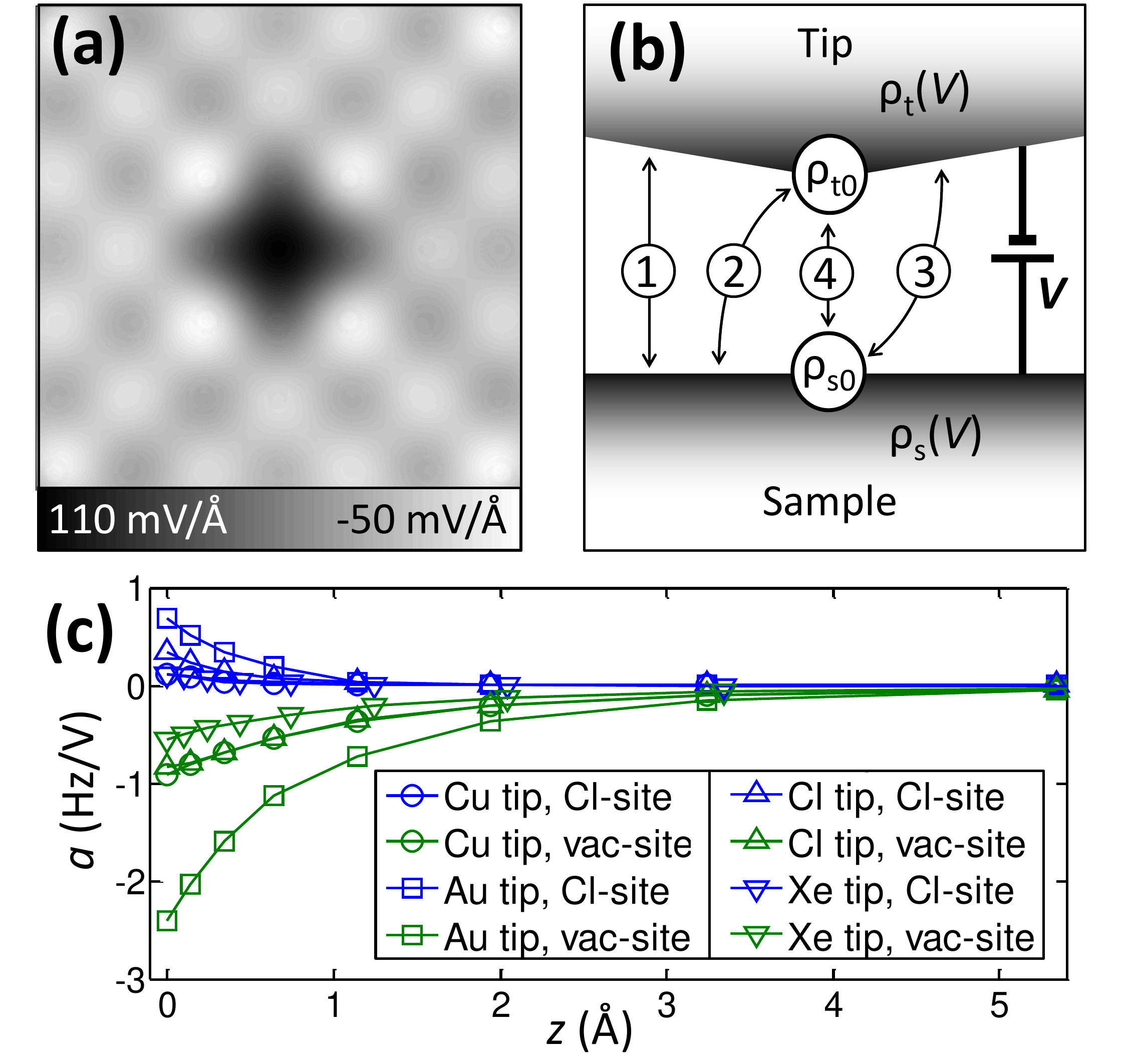}
\caption{\label{fig5}\small (Color online) (a) Calculated $z$-component of the electric field at a distance of 5\,\text{\AA} from the surface plane; image size is $18\,\text{\AA}\,\times\,18\,\text{\AA}$. (b) Schematic showing different contributions of the electrostatic interaction between tip and sample. (c) Linear slope $a(z)$ of $\Delta f(V)$ difference spectra with reference to the Na site.
}
\end{figure}

Using density functional theory (DFT), we calculated the electric field ($E$) above the sample using the PAW code VASP~\cite{Kresse1996,Kresse1999} and employing the optB86b~\cite{Klimes2011} version of the van der Waals density functional. The calculated contrast of the $z$ component of the electric field ($E_{z}$) above the sample shown in Fig.~\ref{fig5}(a) qualitatively agrees with the measured contrast of $V^{*}$ maps [Fig.~\ref{fig4}(a)-(d)]. This agreement can be understood as follows: To compensate for the electric field at the tip position, a matching external field with opposing polarity has to be applied. However, also the mesoscopic parts of tip and sample contribute to the electric field and thus to $V^*$. Because of these aforementioned averaging effects we cannot quantitatively recover the electrostatic potential from $V^*$. 

The atomic tip termination obviously plays an important role for the explanation of the ${\Delta f^*}$ and $F^*_{sr}$ contrast. As we always measured in the regime where ${\Delta f}$ is monotonically decreasing as a function of $z$ and we also see no contrast inversions as a function of $z$, we can rule out significant contributions from Pauli repulsive forces for the investigated tip heights. The contributions from van der Waals forces cannot be responsible for the observed tip-dependent contrast inversions, in particular not for the large attraction above the vacancy for the Au and Cl tip. This leaves the electrostatic interactions as the most important origin of the atomic contrast for our system, in agreement with recent studies by Gao et al.~\cite{Gao2014} and Schneiderbauer et al. ~\cite{Schneiderbauer2014}. 

The schematic in Fig.~\ref{fig5}(b) illustrates different contributions to the electrostatic interaction. The first term is the interaction between the extended homogeneous charge distributions $\rho_t$ and $\rho_s$ of tip and sample, respectively. These charges arise because of the applied $V$ and the different work functions. The 2nd term describes the interaction of a localized, $V$-independent charge distribution $\rho_{t0}$ at the tip with $\rho_s$, and the 3rd term the interaction of a localized, $V$-independent charge distribution $\rho_{s0}$ at the sample with $\rho_t$. Finally, the 4th term describes the interaction of $\rho_{s0}$ with $\rho_{t0}$. Note that the image charges induced by a charge at its corresponding electrode will lead to surface dipoles. These surface dipoles can be considered as areas exhibiting an offset of the local work function with respect to the rest of the electrode.

In a simplified picture of a plate capacitor geometry and local point charges $q_{s0}$ and $q_{t0}$ for the charge distributions $\rho_{s0}$ and $\rho_{t0}$, respectively, we obtain for the electrostatic force:
\begin{equation}
 {F = -(c_1/d)\,(V-V^*_0)^2\,+\,(c_2/d)\,(V-V^*_0)\,q_{t0}}
\nonumber     
\end{equation}
\begin{equation}
{(c_3/d)\,(V^*_0-V)\,q_{s0}\,+\,(c_4/r^2)\,q_{t0}\,q_{s0}}
\end{equation}
where $d$ is the separation of the plates and $r$ is the distance of the charges $q_{s0}$ and $q_{t0}$. Here we neglected polarization of the homogeneously charged plates by $q_{s0}$ and $q_{t0}$. $V^*_0$ is the voltage that compensates the contact potential difference (CPD) of the plates in the absence of $q_{s0}$ and $q_{t0}$. The charge on the plates and the electric field in the junction both increase with $(V-V^*_0)$. The constants $c_1$-$c_4$ are positive and correspond to the four contributions, labeled 1-4 in Fig.~\ref{fig5}(b), respectively.

Next we take into account that the tip has a non-planar shape and that it locally probes the planar sample. Therefore the 2nd term does not depend on the lateral tip position, whereas the 3rd term does. Thus, the 1st and 2nd terms yield a $\Delta f(V)$ parabola that will not depend on the lateral position of the tip. The linear $V$ dependence of the 3rd term is solely responsible for the horizontal shift of the $\Delta f(V)$ parabolas and thus for the $V^*$ contrast. In addition, this term will also contribute to the $\Delta f^*$ contrast because in general, due to aforementioned averaging effects, the electric field at the position of $\rho_{s0}$ is not nullified at $V^*$. Finally, the 4th term, which is $V$-independent, will lead only to a vertical shift of the $\Delta f(V)$ parabolas and thus contribute to the $\Delta f^*$ contrast. 

We subtracted the $\Delta f(V)$ spectra of the Na site as reference from the spectra of the Cl and the vacancy site for a given tip and tip height (see appendix). In this way, we extracted the $V$ dependence of the 3rd term, and indeed the difference spectra are essentially linear confirming experimentally the findings of Sadeghi et al.~\cite{Sadeghi2012} and our simple electrostatic model. We determined the slope $a$ of the difference spectra, shown in Fig.~\ref{fig5}(c). Qualitatively $a(z)$ exhibits a similar behavior for all tips. This experimental finding also supports our simple electrostatic model because the 3rd term is independent of $\rho_{t0}$, i.e., of the tip functionalization. 

To explain the $\Delta f^*$ contrast, we have to consider the 4th term (i.e., the interaction of $\rho_{s0}$ with $\rho_{t0}$). In general the attractive forces that remain at $V^*$ will increase with the inhomogeneity of the electric field in the junction. Opposing tip and sample charges of different sign (or surface dipoles which are parallel) will locally induce a field and thus increase the inhomogeneity of the electric field as compared to opposing charges of the same sign (or anti-parallel dipoles). Localized charges of the sample stem from the ions and the vacancy. The localized charges at the tip, in particular, a tip dipole, arise from the tip shape because of the Smoluchowski effect~\cite{Smoluchowski1941,teobaldi2011} and, additionally, from the tip functionalization. Moreover, also image charges have to be taken into account. The tip dipole moments were calculated with DFT for 10 different tips. The results are shown in Table\,1, with positive $p$ corresponding to a positive partial charge at the tip apex. Two Cu cluster sizes, Cu$_5$ and Cu$_{26}$, were considered to model the tip. For the Au-terminated tips, the foremost Cu tip atoms were replaced by Au atoms. For Cl- and Xe-terminated tips, the corresponding atoms were attached to the foremost Cu tip atom. The tips were fully relaxed while constraining the atoms to keep the tips mirror symmetric with respect to both the (x,z)- and the (y,z)-plane. Table\,1 shows that, except for Au tips, the dipole moments of the smaller five-metal-atom tips agree quite well with those of the larger 26-metal-atom tips. Note that we obtained dipole moments that are comparably large with respect to adsorbates on planar Cu surfaces,~\cite{Sun2011,Silvestrelli2012} which is an effect of the pyramidal shape of the tip apex as shown recently for metallic tips by Gao et al.~\cite{Gao2014}.

\begin{table} [ht]
\begin{center}
\caption{The dipole moments of 10 different tips.}
\begin{tabular}{lrrr} \hline\hline
tip             &  dipole moment $p$ (D)\\ \hline
Cu$_5$          &  0.53\\ 
Cu$_{26}$       &  0.52\\\\[-5mm]
Cu$_4$Au        & -1.05\\
Cu$_{25}$Au     & -0.04\\
Cu$_{23}$Au$_3$ & -1.01\\
Cu$_{21}$Au$_5$ & -2.29\\\\[-5mm]
Cu$_5$Cl        & -5.87\\
Cu$_{26}$Cl     & -7.90\\\\[-5mm]
Cu$_5$Xe        &  2.92\\
Cu$_{26}$Xe     &  3.54\\
\hline\hline
\end{tabular}
\end{center}
\end{table}

The measured $\Delta f^*$ contrast of the Cu, Cl, and Xe tips could be understood from their respective dipole moments: The attraction is increased above sample charges of opposite sign with respect to the tip apex for these three tips. Importantly, the Au tip demonstrates that the $\Delta f^*$ and $F^*_{\text{sr}}$ contrast in general cannot be explained by a tip dipole moment. The Au tip exhibits a larger attraction on both the Cl and the vacancy site than on the Na site, although Cl is charged negatively and the vacancy is charged positively as compared to Na. Two effects could play a role here: (i) The uncompensated positive charge of the vacancy will induce a negative image charge in the tip, thus changing the tip dipole with the tip position. (ii) The charge distributions of tip and sample are more complex than the simple picture of point charges or dipoles. In particular, the lateral charge distribution at the tip due to the tip functionalization (doping) and the Smoluchowski effect should be taken into account. 

\begin{figure} [ht] 
\begin{center}
\includegraphics[width=1\linewidth]{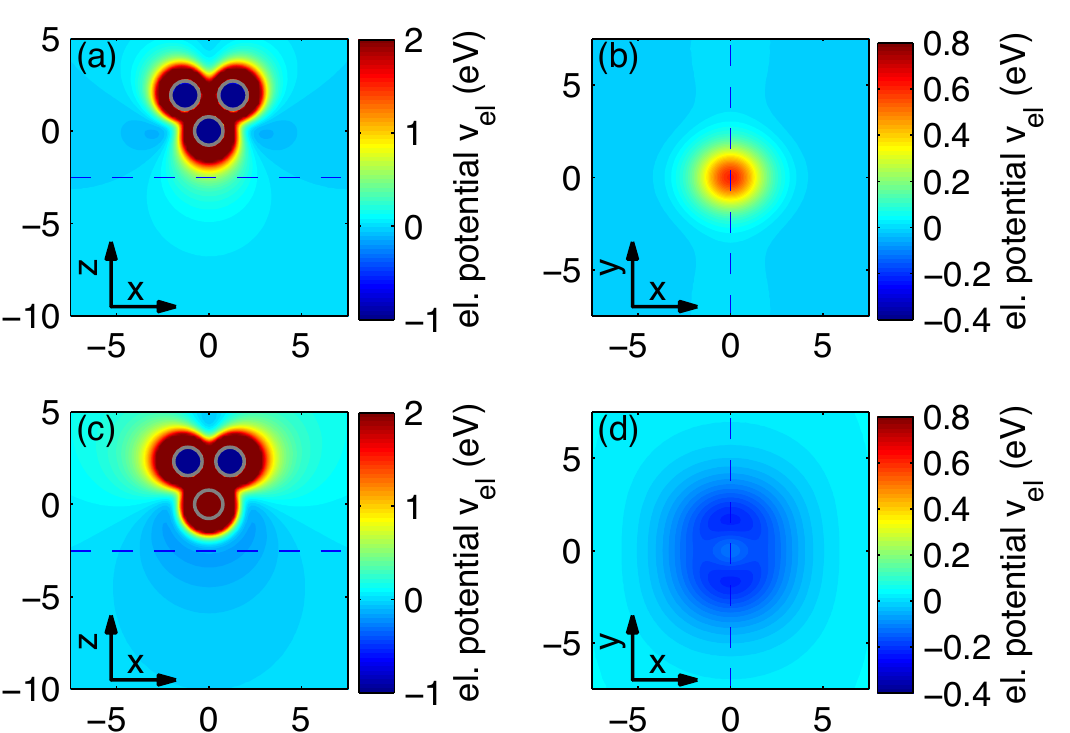}
\caption{\label{fig6}\small The calculated electrostatic potential for a positive elementary charge of the Cu$_5$ tip (a), (b), and of the Cu$_4$Au tip (c), (d). The ($x, z$) planes in (a) and (c) show a vertical cut through the respective tip. The planes (b) and (d) show the potential in the horizontal ($x, y$) plane at a distance of 2.5\,\AA~with respect to the tip atom. The dashed lines in (a) and (c) indicate the planes shown in (b) and (d), and the dashed lines in (b) and (d) indicate the planes shown in (a) and (c), respectively. Note that the four Cu atoms of the second layer of the tip are not in one plane, which is the reason for the two-fold symmetry of the potential maps in (b) and (d).}
\end{center}
\end{figure}

The interplay of the latter effects can be visualized when examining maps of the electrostatic potential of the isolated tips, which are shown for the Cu$_5$ tip in Fig.~\ref{fig6}(a), (b), and for the Cu$_4$Au tip  in Fig.~\ref{fig6}(c), (d). For the Cu$_5$ cluster the potential is increased below the tip atom due to the positive partial charge at the tip apex. For the Cu$_4$Au tip a more complex potential is observed: For small distances the potential is also increased below the tip atom, however, for larger distances the potential below the tip is decreased [blue crescent in Fig.~\ref{fig6}(c)] in correspondence with the long range dipole behavior of the tip. This tip's potential landscape that is exposed towards the sample has a relative maximum located directly below the tip, surrounded by a region of decreased potential energy as shown in~\ref{fig6}(d). 

Such a potential can explain the contrast observed with the Au tip qualitatively. The overall negative potential gives rise to the large attraction measured above the net positively charged vacancy. On NaCl the atomic corrugation of the potential landscape of the tip and of the sample have to be taken into account. Here it is important that the minimum of the tip potential is not located directly below the tip. When the tip is located above a Cl anion, the ring of the potential minima observed in Fig.~\ref{fig6}(d) is positioned above the four neighboring Na cations. This can lead to an increased attraction at the Cl site compared to the Na site, although the tip has a negative dipole moment. The attractive force at compensated LCPD is increased for the tip being above the Na site, because in this case the inhomogeneity of the electric field is increased as compared to the tip being above the Cl site. Importantly, this shows that the simplification of the electric field of the tip by a dipole is not always justified. In particular, the short range field that arises due to the three dimensional charge distribution at the tip apex can be more complex than that of a point dipole and has to be taken into account. Recently, the importance of higher order electrostatic multipoles in the short range field of adsorbed molecules was demonstrated by AFM and KPFM~\cite{Schuler2014}. 

\begin{figure} [ht]
\begin{center}
\includegraphics[width=0.9\linewidth]{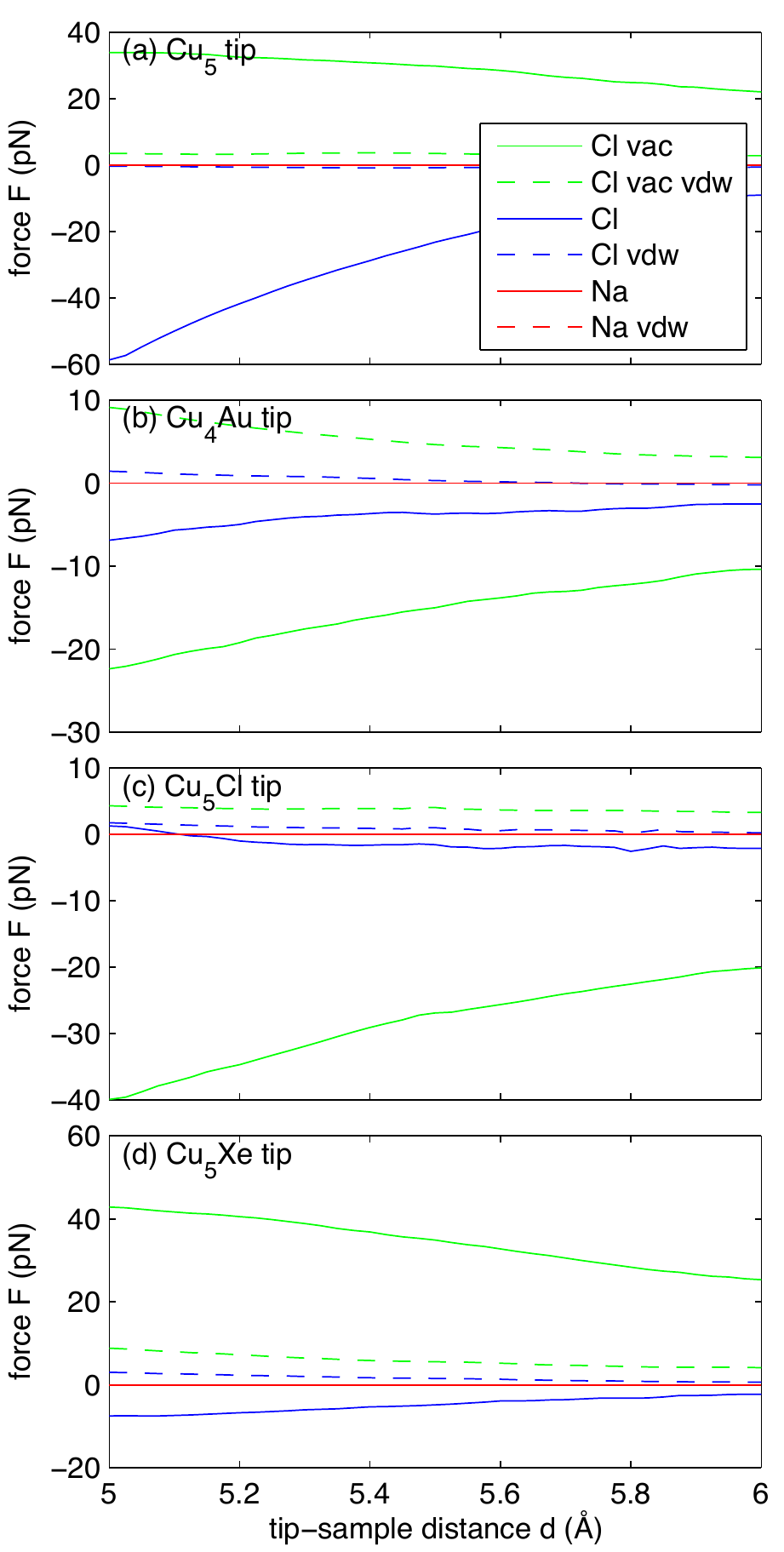}
\caption{\label{fig7}\small The calculated forces $F$ as a function of the tip-sample distance $d$ for four different tips: (a) Cu$_5$ tip, (b) Cu$_4$Au tip, (c) Cu$_5$Cl tip, and (d) Cu$_4$Xe tip above the Cl vacancy, the Cl site, and the Na site (solid lines). The contribution of the van der Waals forces are shown by the dashed lines. All forces are given with respect to the Na site.}
\end{center}
\end{figure}

Next we calculated the interaction forces between the different functionalized Cu$_5$ tips and the sample by DFT using one code with numerical atomic orbitals as basis functions \cite{blum_ab_2009}. The code employed the Perdew-Burke-Ernzerhof exchange-correlation functional (PBE) \cite{perdew_generalized_1996} and a van der Waals method \cite{tkatchenko_accurate_2009} combined with the Lifshitz-Zaremba-Kohn theory for the non-local Coulomb screening within the bulk for the Cu substrate \cite{ruiz_density-functional_2012} using calculated coefficients for {\em atoms in the solid} \cite{zhang_van_2011} for the NaCl film. For the calculations of the tip-sample forces, we used only the five-metal-atom tips in the geometry found by relaxation without the substrate. The forces computed for four different tips and a NaCl surface with a Cl vacancy are shown as a function of the tip height in Fig.~\ref{fig7}. The slab to model the NaCl surface consisted of two layers of Cu substrate with a (001) surface with two layers of NaCl Film on top. The lateral dimensions were $3 \times 3$ for the Cu substrate and $2 \times 2$ for the NaCl film. One Cl vacancy was created, and the NaCl layers were fully relaxed while the Cu substrate was kept fixed. The total energies were calculated for four tips: Cu$_5$, Cu$_4$Au, Cu$_5$Cl, Cu$_5$Xe, and three different lateral positions: above the Cl vacancy, above a Cl atom, and above a Na atom. The atoms of the tips and slab were kept fixed. The tip-sample distances, that is, the distance between the outermost NaCl layer and the center of the tip atom closest to the surface, were varied from 5.0 to 6.0\,\AA, with a spacing of 0.025\,\AA. The forces were obtained by differentiating the total energies numerically. Note that the calculations only take into account the van der Waals interactions for a reduced tip, for example, the Cu$_5$ tip. No macroscopic van der Waals contributions or dipole moments that could arise owing to the macroscopic tip shape were added to the calculations. Therefore attractive forces could be underrepresented in the calculations. However, the forces stemming from the mesoscopic tip shape will not show any corrugation on the atomic scale and therefore will not contribute to the atomic contrast.

When comparing the calculated forces (Fig.~\ref{fig7}) with the experimentally measured ones [Fig.~\ref{fig3}(e)-(h)], we observe an overall good qualitative agreement. The contrast of the vacancy site with respect to the Na and Cl sites is in excellent agreement for all tips: The vacancy appears several 10\,pN more attractive than the Na site with the Au and Cl tips and several 10\,pN less attractive than the Na site with the Cu and Xe tips. The delicate differences between the Na and the Cl site, which amount to only few pN in the experiment, are conformal in magnitude and sign for the calculated Au and Xe tips. For the Cu tip, the sign is also conform, but the difference between the Na and the Cl site is on the order of a few 10\,pN in the calculations, i.e., larger than in the experiment. For the calculated Cl tip, the difference between the Na and the Cl site is only on the order of a few pN, with a height-dependent contrast inversion at $d$\,=\,5.1\,\AA. The experimentally observed contrast between Na and Cl with a Cl tip, i.e., the larger attraction above the Na site, is only observed in the calculations for tip-sample distances smaller than 5.1\,\AA. We also plotted the calculated contributions of the van der Waals forces (dashed lines in Fig.~\ref{fig7}). For all tips, we observe that their contribution to the contrast is relatively small, being about an order of magnitude smaller than the overall force differences. 

In conclusion we obtained the trends correctly by DFT for the contrasts of the vacancy, Cl, and Na sites, but not their magnitudes. The latter might be due to the small tip models used. The calculations showed only minor contributions of van der Waals forces, corroborating that the $\Delta f^*$ contrast arises predominantly from the electrostatic interactions and indicating that the interaction of $\rho_{s0}$ with $\rho_{t0}$ is in main responsible for the $\Delta f^*$ contrast. Note that the electrostatic image charges are included in the DFT calculated interactions. The differences with respect to the simple tip-dipole argument, in particular the theoretical corroboration of the Au tip contrast by DFT, are presumably due to polarization effects~\cite{Kantorovich2000} and/or due to higher order electrostatic multipoles of the tip ($\rho_{t0}$) that are not considered in the simple tip-dipole picture.

\section{IV. Summary and Outlook}

We can now explain the AFM contrast that is obtained on ionic systems as a function of voltage and tip functionalization: $\Delta f^*$ maps reflect the interaction of $\rho_{s0}$ with $\rho_{t0}$ and these maps resemble conventional AFM images acquired at constant voltage of approximately $V^*$ (see Fig.~\ref{fig1}). In this case, the vertical shift of the $\Delta f(V)$ parabolas dominates the contrast because of the small value of the slope of the parabolas near $V^*$. The more the applied voltage deviates from $V^*$, the more will the horizontal shift of the parabolas affect the $\Delta f$ contrast in AFM. The two effects, the horizontal and the vertical shift of the parabolas, can either act together or counteract each other, giving rise to an increase [see Fig.~\ref{fig2}(a)], a decrease [see Fig.~\ref{fig2}(c)] or inversion of atomic contrast with respect to the $\Delta f^*$ maps [Fig.~\ref{fig4}(f)]. Note that in the case of atomically resolved AFM images of molecules using CO tips~\cite{Gross2009a,Mohn2012,Gross2012,Oteyza2013,Zhang2013}, forces of shorter range are explored and the atomic contrast is dominated by Pauli repulsion~\cite{Moll2010,Mohn2012}. In addition, there the interpretation is further complicated by the non-planarity of the sample and the tilting of the CO tip~\cite{Gross2012,Welker2012}. However, also the longer-ranged electrostatic forces that are investigated here will contribute and have to be taken into account for the interpretation of NC-AFM results in general.

Using tip functionalization by atomic manipulation we clarified the properties and the origin of AFM and KPFM contrast on the atomic scale: We found almost no tip dependence of the relative contrast in the $V^*$ (LCPD) measured with KPFM, but very pronounced tip- and voltage-dependent contrast for AFM. Remarkably, electrostatic forces are the main contributions in all cases. The $V^{*}$ channel reflects the $z$-component of the electric field ($E_{z}$) above the sample. The AFM contrast at compensated LCPD depends crucially on the tip because of the direct electrostatic interaction of localized, voltage-independent tip and sample charges. The complementary properties of the ${\Delta f^*}$ and the $V^*$ channels could be very beneficial: Maps of ${\Delta f^*}$ acquired with different tips can be related to each other using the tip-independent $V^*$ maps as reference, thus facilitating chemical fingerprinting of both the tip and the sample.

\begin{acknowledgments}
\section{Acknowledgments}
The authors thank J. Repp and R. Allenspach for valuable comments, and the EU projects ARTIST (contract no. 243421), PAMS (contract no. 610446), QTea and the ERC advanced grant CEMAS for financial support. Allocation of computer resources at HECToR through the MCC funded by EPSRC (EP/L000202F) and at Lindgren, PDC through SNIC is also acknowledged.
\end{acknowledgments}

\appendix
\section{Appendix A: Complete Data Sets and Determination of Linear Slope} 
\setlength{\parskip}{0pt}
The complete datasets of the four different tips are shown in Fig.\,A1(a): Cu tip,  Fig.\,A2(a): Au tip, Fig.\,A3(a): Cl tip, and Fig.\,A4(a): Xe tip. For each tip, the $\Delta f(V)$ spectra above the different sites (Na, Cl, and vac) are shown for the different tip heights $z$ measured. For the Cu, Au, and Cl tips eight different tip heights were measured: $z$\,=\,[0\,\AA, 0.14\,\AA, 0.34\,\AA, 0.64\,\AA, 1.14\,\AA, 1.94\,\AA, 3.24\,\AA, 5.34\,\AA]. For the Xe tip nine different tip heights were measured: $z$\,=\,[0\,\AA, 0.1\,\AA, 0.24\,\AA, 0.44\,\AA, 0.74\,\AA, 1.24\,\AA, 2.04\,\AA, 3.34\,\AA, 5.44\,\AA]. For a certain tip and site, the values of $\Delta f(V)$ are always increasing with increasing $z$. The $\Delta f(V)$ parabolas exhibiting the most negative $\Delta f$ values correspond to $z$\,=\,0\,\AA. In Figs.\,A1 to A4, panels (b) and (c) correspond to the difference spectra with respect to the Na site for (b) the Cl site and (c) the vacancy (vac) site. The linear fits are shown as red dashed lines. It can be seen that for all tips the difference spectra are well approximated by linear fits. The slope $a(z)$ of these linear fits is plotted in the respective panel (d); the lines connecting the data points are drawn as a guide to the eye.

\makeatletter 
\setcounter{figure}{0} 
\renewcommand{\thefigure}{A\@arabic\c@figure}
\makeatother
\begin{figure*} 
\begin{center}
\includegraphics[width=16cm]{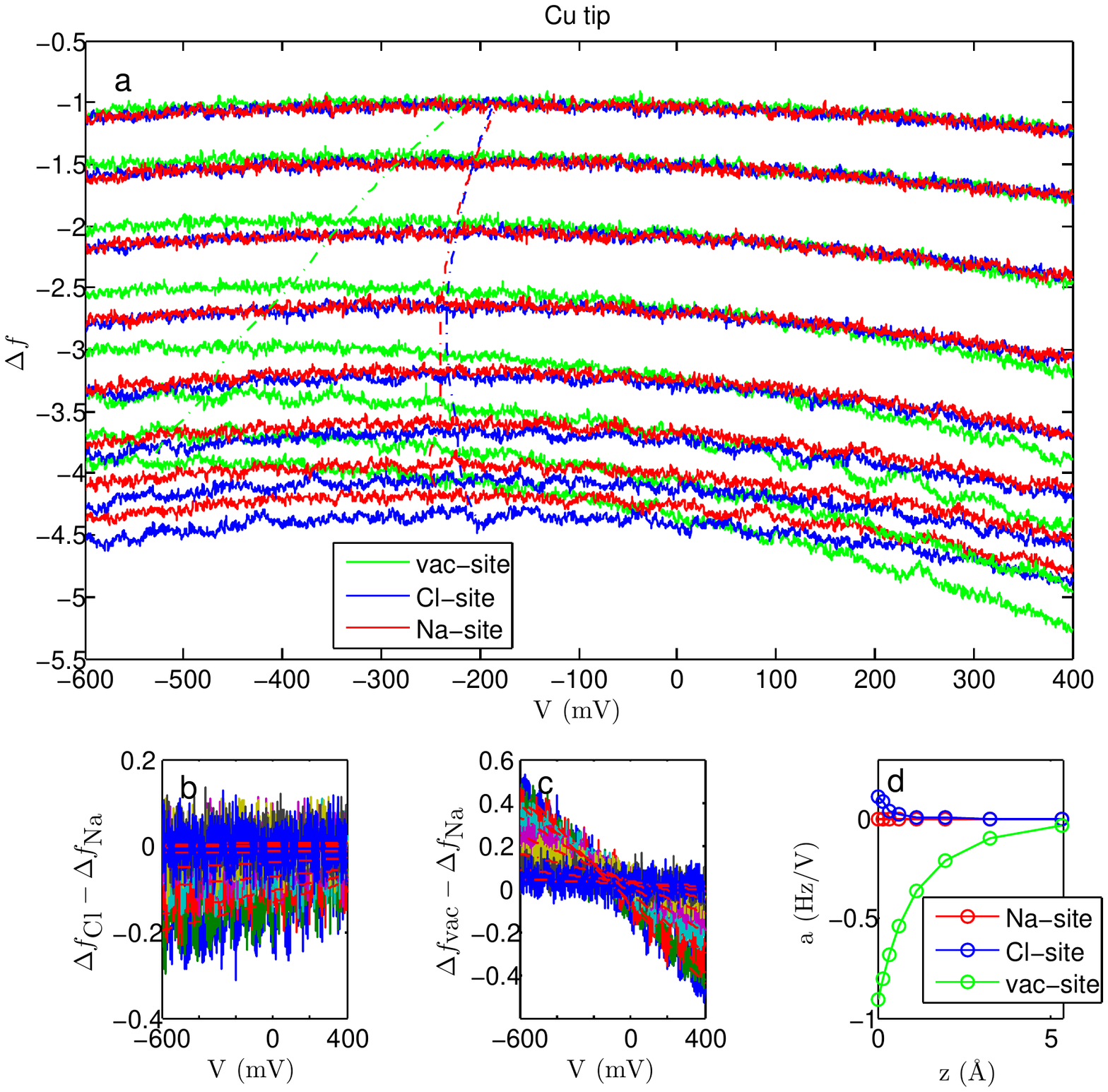}
\caption{(a) Cu tip raw data ($\Delta f(V)$ spectra) at different sites and different tip heights, with the evolution of the peaks ($V^*, \Delta f^*$) indicated by the dashed lines. Difference spectra with respect to the Na site for (b) the Cl site and (c) the vac site. (d) Linear slope $a(z)$.}
\end{center}
\end{figure*}

\begin{figure*} 
\begin{center}
\includegraphics[width=16cm]{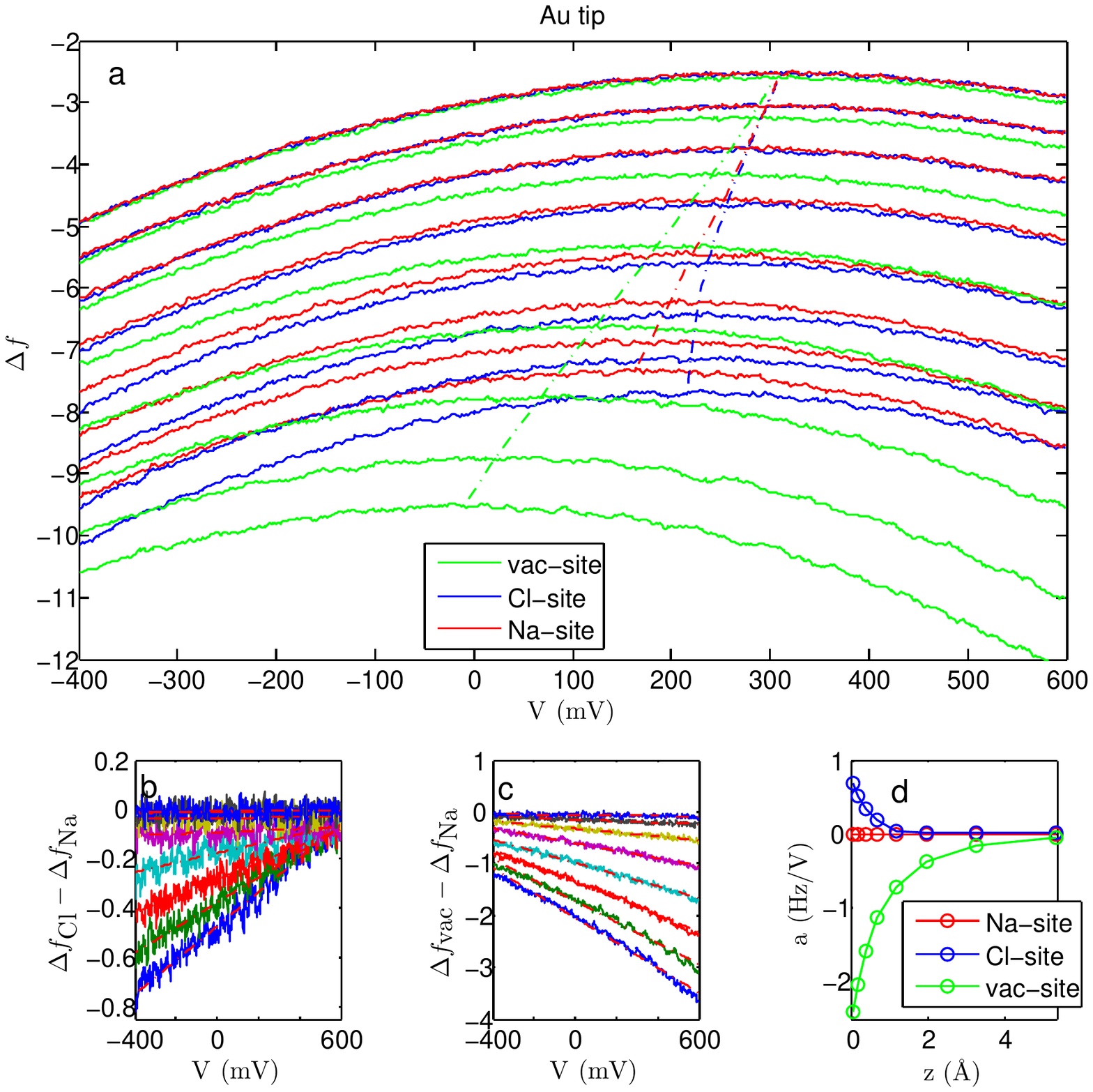}
\caption{(a) Au tip raw data ($\Delta f(V)$ spectra) at different sites and different tip heights, with the evolution of the peaks ($V^*, \Delta f^*$) indicated by the dashed lines. Difference spectra with respect to the Na site for (b) the Cl site and (c) the vac site. (d) Linear slope $a(z)$.}
\end{center}
\end{figure*}

\begin{figure*}
\begin{center}
\includegraphics[width=16cm]{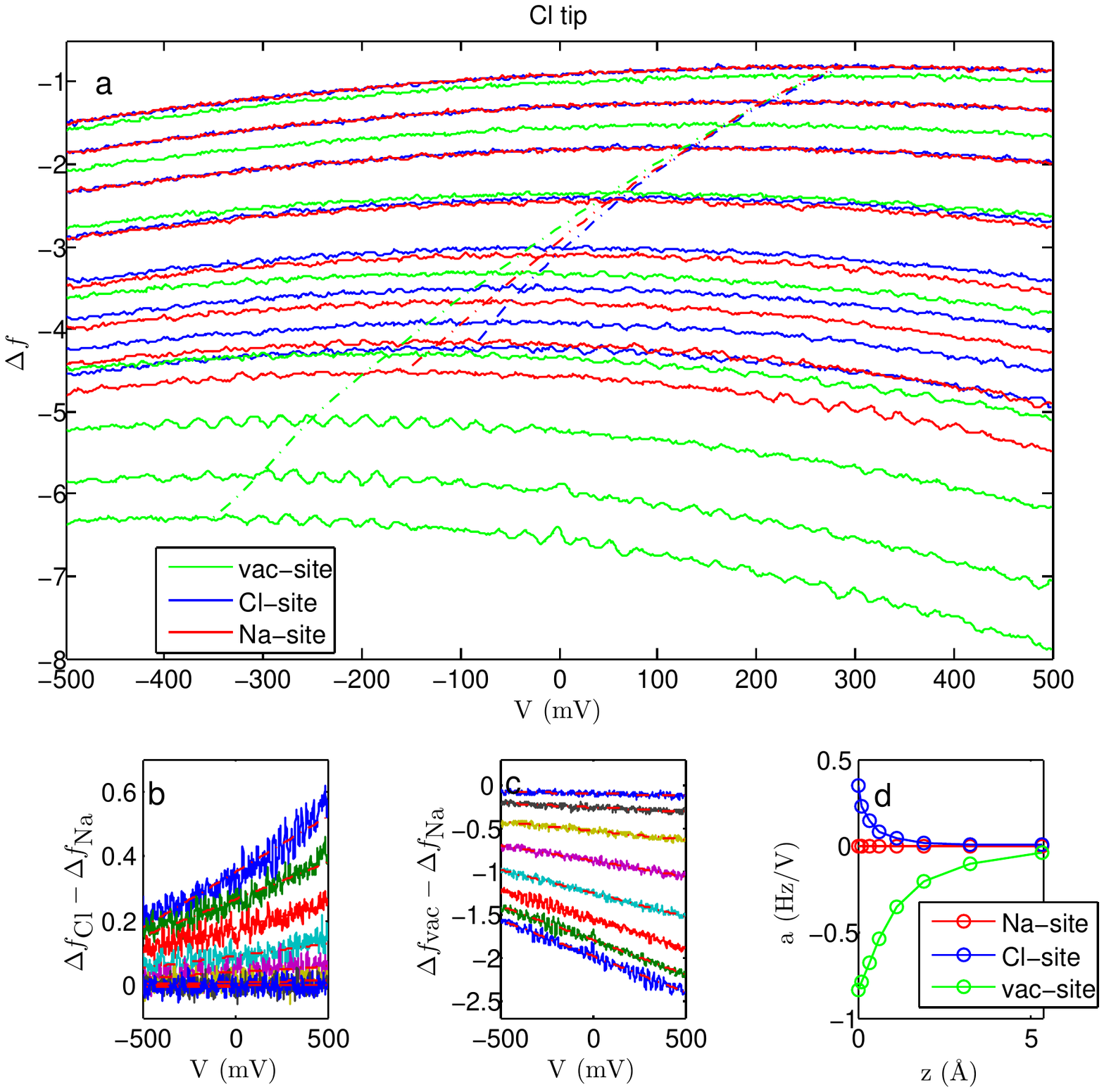} 
\caption{(a) Cl tip raw data ($\Delta f(V)$ spectra) at different sites and different tip heights, with the evolution of the peaks ($V^*, \Delta f^*$) indicated by the dashed lines. Difference spectra with respect to the Na site for (b) the Cl site and (c) the vac site. (d) Linear slope $a(z)$.}
\end{center}
\end{figure*}

\begin{figure*}
\begin{center}
\includegraphics[width=16cm]{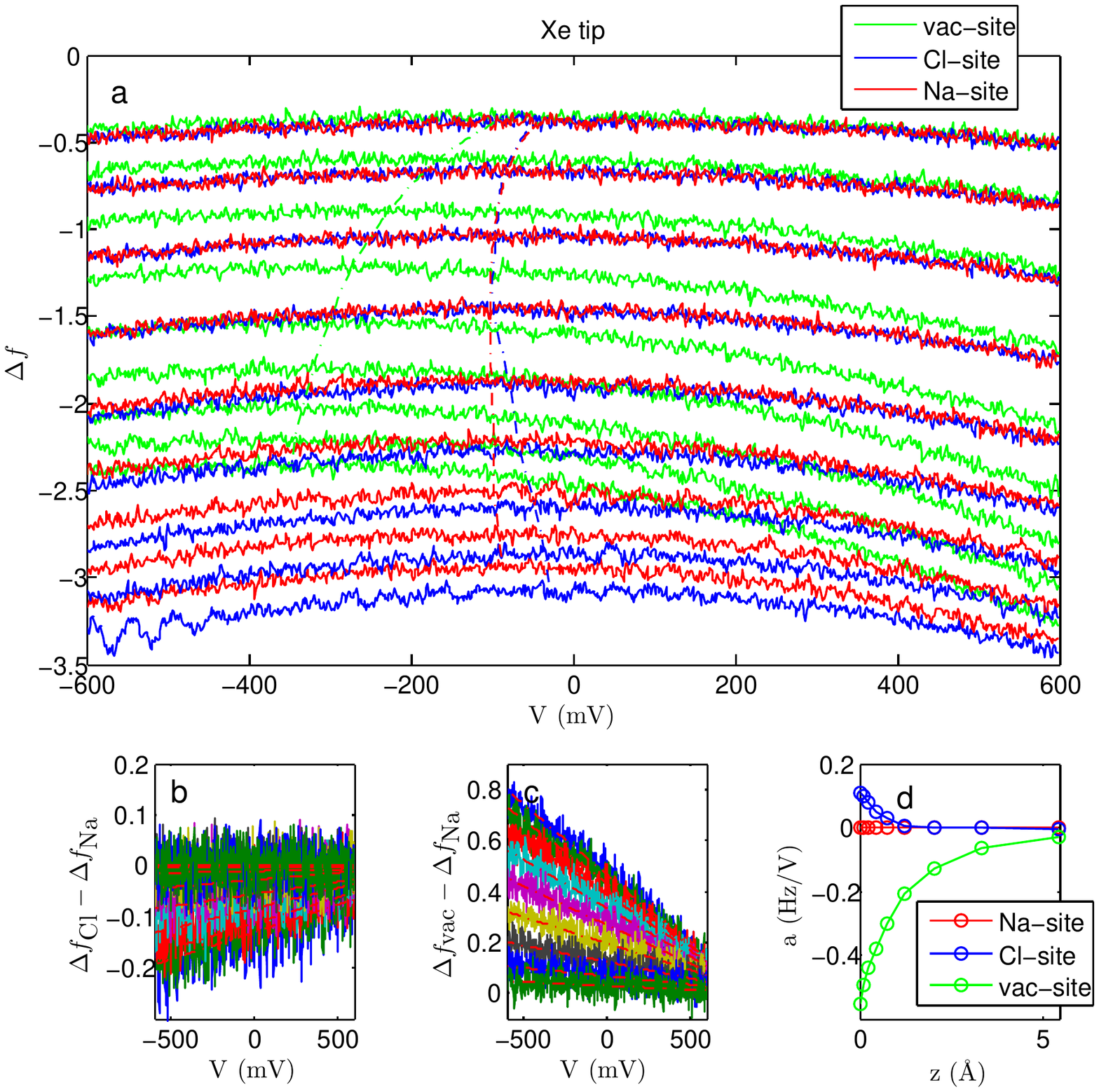} 
\caption{Xe tip raw data ($\Delta f(V)$ spectra) at different sites and different tip heights, with the evolution of the peaks ($V^*, \Delta f^*$) indicated by the dashed lines. Difference spectra with respect to the Na site for (b) the Cl site and (c) the vac site. (d) Linear slope $a(z)$.}
\end{center}
\end{figure*}

\appendix
\section{Appendix B: Atom-by-Atom Fabrication of an Au Tip}

\setlength{\parskip}{0pt}
To corroborate that it is indeed the atomic termination that governs the contrast of a tip, we measured AFM images of a Cl vacancy before [Fig.\,A5(a)] and after [Fig.\,A5(c)] we picked up individual Au adatoms with a Cu tip to form an Au tip. We started with a sharp, symmetric Cu tip and imaged a Cl vacancy at $V$\,=\,0\,V using AFM, obtaining the typical image of a Cu tip [bright vacancy, Cl darker than Na, compare with Fig.~\ref{fig1}(a)]. Then we picked up single Au adatoms from double-layer NaCl/Cu(111) and imaged the same vacancy with the resulting tip again. When we picked up the first and the second Au adatom, the resulting tip did not resolve the vacancy symmetrically, and apparently an atomic double tip was created in these events. An image of the vacancy after two Au adatoms had been picked up is shown in Fig.\,A5(b). Finally, after picking up the third Au adatom, a sharp symmetric tip was created again, showing a different contrast than the initial Cu tip [compare Fig.\,A5(a) and Fig.\,A5(c)]. This we assigned as an Au tip and it exhibited the characteristic Au-tip AFM contrast [dark vacancy, Cl darker than Na, compare with Fig.~\ref{fig2}(b) or Fig.~\ref{fig1}(b)]. Note that the tips used in Fig.\,A5 are different ones than the Cu tip and the Au tip characterized in the main manuscript. 

\begin{figure}
\begin{center}
\includegraphics[width=1\linewidth]{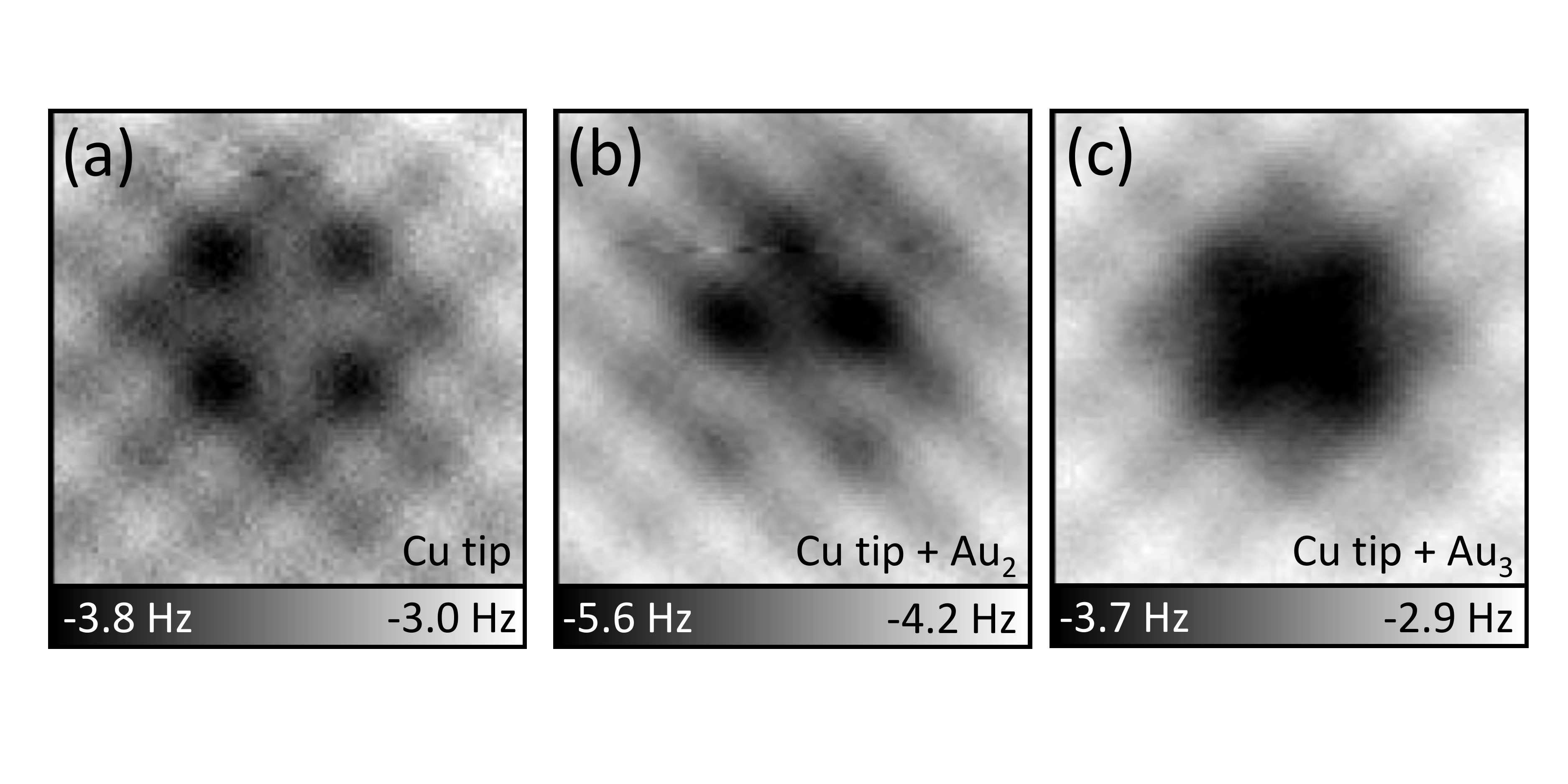}
\caption{Constant-height, constant-voltage ($V$\,=\,0\,mV) AFM measurements of a Cl vacancy using (a) a Cu tip, (b) the same Cu tip after picking up 2 individual Au adatoms, and (c) after picking up a 3rd individual Au adatom. Note that a typical Cu tip AFM image of the vacancy is obtained in (a) and a typical Au tip AFM image in (c).}
\end{center}
\end{figure}

Usually several Au adatoms had been picked up with a Cu tip to create a sharp symmetric Au tip and these tips showed the typical contrast of an Au tip. (For the Xe and the Cl tips only a single atom was picked up.) Often the pickup of a single Au adatom with a very sharp atomic Cu tip resulted in a less symmetric and presumably less sharp tip. The positions of the Au atoms that are incorporated into the tip before picking up the final, terminating Au atom are not known. As the images of the sharp Au tip made by picking up several Au adatoms are symmetric, we assume that the foremost Au atom of the tip is crucial for the contrast and the other incorporated Au atoms play only a minor role. However, our calculations (see Table.\,1) show that the overall tip dipole moment changes with the number of Au tip atoms. Therefore we expect that Au tips with different dipole moments could be created, which could be used to fabricate tips with customized dipole moments. We did not investigate the KPFM contrast in dependence of the number of Au atoms picked up. Importantly, the Au tip investigated in the manuscript demonstrates that tips exist, whose $\Delta f^*$ contrast cannot be understood from the dipole moment of the isolated tip.


%

\end{document}